\def\bar{\begin{eqnarray}}
\def\ear{\end{eqnarray}}

\def\eqi{\begin{equation}}
\def\eqf{\end{equation}}
\def\eqia{\begin{eqnarray}}
\def\eqfa{\end{eqnarray}}
\def\rp#1#2{{#1\over#2}}

\def\oc2{$\mathcal{O}(c^{-2})$}

\documentclass[11pt]{article}
\usepackage{amsmath,amsthm,amscd,amssymb}
\usepackage{latexsym}
\usepackage{graphicx,epsfig}
\usepackage[edtable]{lineno}

\begin{document}
%\pagewiselinenumbers % ad ogni pagina riprende da 1 il conteggio delle linee
%\runninglinenumbers % conta le linee da 1 a X fino alla fine del file

\noindent{\bf \LARGE{A comment on the paper ``On the orbit of the
LARES satellite'', by I. Ciufolini}}
\\
\\
\\
{Lorenzo Iorio}\\
{\it Viale Unit$\grave{a}$ di Italia 68, 70125\\Bari, Italy
\\tel./fax 0039 080 5443144
\\e-mail: lorenzo.iorio@libero.it}

\begin{abstract}
In this note we comment on a recent paper by I.Ciufolini about the
possibility of placing the proposed terrestrial satellite LARES in
a low-altitude, nearly polar orbit in order to measure the general
relativistic Lense-Thirring effect with its node. Ciufolini claims
that, for a departure of 4 deg in the satellite's inclination $i$
from the ideal polar configuration ($i=90$ deg), the impact of the
errors in the even zonal harmonics of the geopotential, modelled
with EIGEN-GRACE02S, would be nearly zero allowing for a
few-percent measurement of the Lense-Thirring effect. Instead, we
find that, with the same Earth gravity model and for the same
values of the inclination, the upper bound of the systematic error
due to the mismodelling in the even zonals amounts to 64$\%$ of
the relativistic effect investigated.
\end{abstract}


\section{The polar configuration for measuring the Lense-Thirring effect}
The possibility of measuring the general relativistic
gravitomagnetic Lense-Thirring effect by means of the node of a
LAGEOS-like satellite placed in a relatively low-altitude ($a\sim
8000$ km), polar ($i\sim 90$ deg) orbit$-$POLARES in the
following$-$was proposed for the first time by Lucchesi and
Paolozzi (2001) and, subsequently, criticized by Iorio (2002). The
benefits of such an idea mainly rely in the possibility of using a
relatively cheap launcher vehicle and in the fact that, for a
perfectly polar configuration ($i=90$ deg), all the classical
secular precessions induced on the node by the even
($\ell=2,4,6...$) zonal ($m=0$) harmonic coefficients $J_{\ell}$
of the Newtonian multipolar expansion of the terrestrial
gravitational potential, proportional to $\cos i$, vanish. The
main drawbacks of such an orbital configuration are as follows
\begin{itemize}
\item The satellite's node is perturbed, among other things, by
the $\ell=2, m=1$ constituent of the Solar $K_1$ tide whose period
is equal to that of the spacecraft's node itself: for $i\sim 90$
deg it precesses very slowly, so that $K_1$ would mimic an
aliasing secular trend over an observational time span of a few
years compromising the recovery of the genuine relativistic linear
trend of interest. This general feature of motion of a polar
satellite was already recognized by Peterson (1997) and Iorio
(2005a) in the framework of the GP-B mission.
\item
This problem is avoided by choosing an inclination a few deg apart
from the ideal polar configuration. But, in this case, the
systematic error $\delta\mu$ induced by the mismodelled part of
all the even zonal harmonics, emphasized by the low altitude of
the satellite, is enhanced, depending on the accuracy of the
gravity model used. Iorio (2002) used the full covariance matrix
of EGM96 (Lemoine et al. 1998) up to degree $\ell=20$ showing that
for orbits just 1 deg apart from $i=90$ deg the impact of the
mismodelled even zonals considered amounted to 40$\%$. Such an
estimate is likely optimistic because of the use of the
correlations among the solved-for even zonals.
\end{itemize}
In (Iorio 2002) the possibility of using POLARES in conjunction
with the existing LAGEOS and LAGEOS II satellites according to the
well-known linear combination approach was investigated as well:
it turned out to be unfeasible because of the quite large value of
the coefficient with which the POLARES node would enter such
combinations.

Iorio (2005b) extensively studied the impact of the new Earth
gravity models by CHAMP and GRACE on the possibility of using a
new satellite to measure the Lense-Thirring effect. Among other
things, the POLARES configuration ($a=8000$ km and $e=0.04$) was
re-analyzed with the EIGEN-CG01C model (Reigber et al., 2006), up
to degree $\ell=20$ and, much more conservatively than in (Iorio
2002), by linearly summing up the absolute values of the
individual mismodelled classical precessions; the situation is now
improved with respect to the EGM96 case, but it turned out that
for a shift of just 2 deg in $i$ with respect to the ideal polar
geometry the bias due to the mismodelling in all the uncancelled
even zonal harmonics still amounts to about 25$\%$. In Iorio
(2005b) also the linear combination scenario with LAGEOS and
LAGEOS II was investigated showing that for $a=8000$ km, $e=0.04$,
and 60 deg$<i<$80 deg the systematic error due to the mismodelled
even zonals is 1-3$\%$.

\section{The departures from the ideal polar configuration}
The subject seemed, thus, to have exhaustively been treated so
far, when a new paper on it by Ciufolini (2006)  appeared.
Basically, the only novelty of such work, which mainly reproduces
the content of Section 4.2.1 and Section 4.4 of Iorio (2005b)
without quoting it, is a huge underestimation of the impact of the
uncertainties in our knowledge of the geopotential on a certain
orbital configuration of POLARES. Indeed, Ciufolini (2006), who
used the EIGEN-GRACE02S Earth gravity model (Reigber et al.,
2005), after discussing the problem of the $K_1$ tide, proposed to
circumvent it by adopting for POLARES an orbital configuration
with $a=7878$ km and $i\leq 86$ deg or $i\geq 94$ deg, i.e. an
inclination's  departure of 4 deg from the ideal polar geometry.
He explicitly claims that, in this case ``[...] if LARES would be
launched in a nearly polar orbit the use of LAGEOS and LAGEOS 2
satellites would not be anymore useful in order to reduce the
error budget (and would indeed only introduce an additional
error), since the effect of the even zonal harmonics on the node
of LARES would be nearly zero, [...]''. Unfortunately, the
situation is quite different, as it could already have been
inferred from Section 4.2.1 and Figure 5 of (Iorio 2005b). For the
sake of a direct comparison, here we use EIGEN-GRACE02S as well,
and in order to get a conservative upper bound of the systematic
error induced by the mismodelling in $J_{\ell}$ we linearly sum up
the absolute values of the individual mismodelled node precessions
up to $\ell=40$ \eqi\delta\mu\leq \sum_{\ell
=2}^{40}\left|\dot\Omega_{.\ell}\right|\delta J_{\ell}\eqf by
using the calibrated errors in $J_{\ell}$ (Reigber et al. 2005).
The coefficients $\dot\Omega_{.\ell}$ of the classical node
precessions were explicitly worked out  up to degree $\ell=20$ in
(Iorio 2003): for, e.g., $\ell=2$ we have
\eqi\dot\Omega_{.2}=-\rp{3}{2}n\left(\rp{R}{a}\right)^2\rp{\cos i
}{(1-e^2)^2},\eqf where $R$ denotes the Earth's mean equatorial
radius and $n=\sqrt{GM/a^3}$ is the Keplerian mean motion.
Ciufolini (2006) did not explain how he assessed the error due to
the even zonals (root-sum-square calculation? Sum of the absolute
values of the individual errors?), apart from claiming that he
used the analytical expressions of the nodal precession of a
satellite, up to $\ell=10$, from an unspecified reference R.
Tauraso (2004). For $a=7878$ km, $e=0.04$ and $i=86/94$ deg we
get\footnote{If we truncate the calculation at $\ell=20$ and
$\ell=10$ we get $\delta\mu\leq 63\%$ and $\delta\mu\leq 59\%$,
respectively.} \eqi\delta\mu\leq 64\%.\eqf In Table \ref{smerd} we
release the details of our calculation. As can be noted, the
uncancelled precession due to $\delta J_2$ amounts to 70$\%$ of
the entire error.
{\small\begin{table}\caption{Individual mismodelled node
precessions
$\delta\dot\Omega^{(J_{\ell})}\equiv\left|\dot\Omega_{.\ell}\right|\delta
J_{\ell}$ induced by the calibrated errors  in $J_{\ell},\
\ell=2,4,6...40$, in milliarcseconds per year (mas yr$^{-1}$),
according to the variance matrix of EIGEN-GRACE02S Earth gravity
model (Reigber et al. 2005) for $a=7878$ km, $e=0.04$, $i=86$ deg.
The mismodelled precessions for $\ell\geq 30$ are smaller than 0.1
mas yr$^{-1}$. The Lense-Thirring effect for such an orbital
configuration amounts to 116.6 mas yr$^{-1}$. The upper bound of
the total error $\delta\mu\leq
\sum_{\ell=2}^{40}\delta\dot\Omega^{(J_{\ell})}$ is quoted, in mas
yr$^{-1}$, in the last row: it amounts to 64$\%$ of the
Lense-Thirring effect. The most important contribution comes from
$J_2$ whose mismodelled precession amounts to 70$\%$ of the total
error. }\label{smerd}

\begin{tabular}{lllll}
\noalign{\hrule height 1.5pt}

Degree $\ell$ & $\delta\dot\Omega^{(J_{\ell})}$ (mas yr$^{-1}$)\\
\hline
2& 47.9\\
4& 5.8\\
6& 3.4\\
8& 2.4\\
10& 3.0\\
12& 1.5\\
14& 1.1\\
16& 0.7\\
18& 0.5\\
20& 0.4\\
22& 0.3\\
24& 0.2\\
26& 0.1\\
28& 0.1\\
30& $-$\\
32& $-$\\
34& $-$\\
36& $-$\\
38& $-$\\
40 & $-$\\

\hline
$\delta\mu$ & 68.1 \\

\noalign{\hrule height 1.5pt}
\end{tabular}

\end{table}}

In addition to the static part of the geopotential, also its
time-dependent components must also be considered. In particular,
for $i=90\pm 4$ deg, the mismodelled part of the $\ell=2$ $m=0$
constituent of the 18.6-year tide would have a serious aliasing
impact on a sought few-percent measurement, especially over an
observational time span of just 3 years, as proposed by Ciufolini
(2006). The uncancelled secular variations of the even zonals
$\dot J_2,\dot J_4, \dot J_6$ would be another source of
systematic error.

Thus, it seems to us very difficult to agree with the conclusion
by Ciufolini (2006) ``A nearly polar orbit for LARES at an
altitude of about 1500 km would be suitable for a measurement of
the Lense-Thirring effect with accuracy of a few percent.''. A new
satellite can be fruitfully used only in conjunction with  LAGEOS
and LAGEOS II. Such existing satellites, however, would set the
total realistic accuracy obtainable to a few percent level because
of the impact of the non-gravitational forces acting on them,
independently of how well they could be reduced on LARES. Indeed,
it is not clear if and how LAGEOS and LAGEOS II could benefit from
the reduction of the non-gravitational forces on LARES. Such
interesting technological and engineering efforts (Bellettini et
al. 2006) could likely turn out to be really and fully useful  if
the launch of at least two entirely new spacecraft was implemented
(Iorio 2005b).

%-------------------------------------

\end{document}